\documentclass{article}
\usepackage{fullpage}
\usepackage{PRIMEarxiv}
\usepackage[hyphens]{url} 
\usepackage{graphicx} 
\usepackage[utf8]{inputenc} 
\usepackage[T1]{fontenc}    
\usepackage{hyperref}       
\usepackage{url}            
\usepackage{booktabs}       
\usepackage{amsfonts}       
\usepackage{nicefrac}       
\usepackage{microtype}      
\usepackage{lipsum}
\usepackage{fancyhdr}       
\usepackage{graphicx}       
\graphicspath{{media/}}     
\usepackage{natbib}
\usepackage{times} 
\usepackage{helvet} 
\usepackage{courier} 
\usepackage{natbib} 
\usepackage{caption} 
\usepackage{algorithm}
\usepackage{algorithmic}
\usepackage{multirow}
\usepackage{multicol}
\usepackage{amsmath}
\usepackage{booktabs}
\usepackage{subcaption}

\usepackage{enumitem}

\usepackage{newfloat}
\usepackage{listings}
\usepackage{tikz}
\DeclareCaptionStyle{ruled}{labelfont=normalfont,labelsep=colon,strut=off} 
\floatstyle{ruled}
\newfloat{listing}{tb}{lst}{}
\floatname{listing}{Listing}

\usepackage{tikz}
\usetikzlibrary{shapes.geometric, arrows}
\newcommand{\step}[2]{\noindent\paragraph{Step #1: #2}}
\title{(Un)fair Mistakes on Social Media: How Demographic Characteristics influence Authorship Attribution}

\author{%
  Jasmin Wyss\\ 
        \textit{Ruhr University Bochum} \\\textit{jasmin.wyss@rub.de}
  \and Rebekah Overdorf\\ 
        \textit{Ruhr University Bochum}\\
        \textit{Research Center Trustworthy Data Science}\\
        \textit{and Security in University Alliance Ruhr}\\
        \textit{rebekah.overdorf@rub.de}
}

\begin{document}
\maketitle
\begin{abstract}
Authorship attribution techniques are increasingly being used in online contexts such as sock puppet detection, malicious account linking, and cross-platform account linking. Yet, it is unknown whether these models perform equitably across different demographic groups. Bias in such techniques could lead to false accusations, account banning, and privacy violations disproportionately impacting users from certain demographics. In this paper, we systematically audit authorship attribution for bias with respect to gender, native language, and age. We evaluate fairness in 3 ways. First, we evaluate how the proportion of users with a certain demographic characteristic impacts the overall classifier performance. Second, we evaluate if a user's demographic characteristics influence the probability that their texts are misclassified. Our analysis indicates that authorship attribution does not demonstrate bias across demographic groups in the closed-world setting. Third, we evaluate the types of errors that occur when the true author is removed from the suspect set, thereby forcing the classifier to choose an incorrect author. Unlike the first two settings, this analysis demonstrates a tendency to attribute authorship to users who share the same demographic characteristic as the true author. Crucially, these errors do not only include texts that deviate from a user's usual style, but also those that are very close to the author's average. Our results highlight that though a model may appear fair in the closed-world setting for a performant classifier, this does not guarantee fairness when errors are inevitable. 
\end{abstract}

\begin{multicols}{2}

\section{Introduction}

Pseudonymous social networks, such as Reddit, Discord, and Telegram, differ from social networks like Facebook in that the majority of users do not display their real names. Reddit\footnote{\url{https://redditinc.com/privacy}} and Discord\footnote{\url{https://discord.com/safety/identity-authenticity-policy-explainer}} explicitly allow for accounts with no real name associated with them, and Telegram recently updated its account creation protocol to allow for anonymous accounts~\cite{Umar_Shakir_2025_theverge}. Because no tie to a real-world identity is required to be publicly visible, users on these platforms have some expectation of anonymity.

Authorship attribution, an application of stylometry, undermines the pseudonymous protections that users rely on for privacy and anonymity. In the same way that users can leverage anonymity for malicious (e.g., harassment, disinformation) or constructive (e.g., peer support on sensitive topics, whistleblowing) purposes, authorship attribution can be applied in both harmful and protective ways. Platforms or moderators can use stylometry to strengthen platform safety and enforce platform policies and rules by finding sock puppet accounts~\cite{sockpuppetswikipedia}, detecting coordinated malicious accounts~\cite{telegrambots}, and linking malicious users across platforms~\cite{afroz2014doppelganger}. However, stylometry can also be used to deanonymise genuine users~\cite{9424712}. For example, anonymous comments about users' own negative experiences working for a company or organisation are common on Reddit and are often voted to the top of comment threads. The target of such a comment could use stylometry to determine which of their employees authored the comment. Similarly, because Reddit posts can contain sensitive information~\cite{brown2018reddit}, another user who suspects that the author of such a post is someone they know could employ stylometry to confirm their suspicion. 

This dual-use nature of stylometry makes it especially important to study its biases and shortcomings. When used maliciously, understanding the weaknesses can help in building better defences to protect user anonymity. When used for protective purposes, understanding the weaknesses helps us in better interpreting the results. For example, if the sockpuppet detection mechanism is prone to falsely identifying accounts based on topic-similarity, this would need to be mitigated and addressed in the interpretation of the results. Especially in cases where the classes are imbalanced, good performance metrics, such as high F1 or high accuracy, are not enough. It is crucial to test the mechanisms for biases and third variable influences. 
We draw from the structures of pseudonymous social networks and design three approaches to test if there is bias in authorship attribution. 

First, we vary the proportion of users from different demographic backgrounds in the training data and evaluate the impact on the classification score. This reflects the real-world conditions on social networks where users are not necessarily evenly distributed by their demographic traits. If, e.g., a high percentage of non-native English speakers in the suspect set negatively impacts classification performance, it would follow that a gaming channel on Discord dominated by users from Southeast Asia would have a lower detection rate for sockpuppet accounts. On Reddit, subreddits like r/europe, where many users are not native English speakers but most posts and comments are in English, would be similarly impacted. 

Second, we examine who is impacted by the errors. For this, we compare the misclassification rates across different demographic groups. Consider a subreddit focused on domestic abuse where most of the users are women. If women are more likely to be correctly classified than men, this would cause more women to be subject to successful deanonymization attacks on their sensitive posts. 

Finally, we evaluate if demographic characteristics influence the errors made when the true author is removed from the suspect set, thereby forcing the classifier to choose an incorrect author. This allows for an examination of the influence of demographic characteristics on writing style without measuring the consistency of a user's writing style. Users do not always write using a consistent style (e.g., emotional posts, new topics). As such, many of the comments that are misclassified in the previous methods are samples in which the author's style diverges most from their normal style. 
 
\subsection{Contributions}

\begin{itemize}[leftmargin=*]
\item We systematically evaluate the fairness of authorship attribution and show that: 
\begin{itemize}
    \item The demographic makeup of a set does not impact the overall classifier performance. 
    \item The demographic characteristics of a user do not impact the probability that their text is misclassified.
    \item In a forced classification setting where the true author is not present, the demographic characteristics \emph{do} impact who is falsely accused. 
\end{itemize}
\item We present a new methodology, \textit{forced misclassification}, to systematically evaluate which users are falsely accused based on their demographic characteristics when the true author is not present in the training data. 
\item Further, we adapt fairness metrics from the literature into the setting of authorship attribution on social media.
\item We present a novel data collection methodology that combines the Reddit and Wayback Machine APIs for mining relevant Reddit threads. Using this combined approach, we were able to collect at most 2,071\% more threads versus using the Reddit API alone.

\end{itemize}

\section{Related Work}
\paragraph{Authorship Attribution or Verification}

Stylometry applied to text from multiple accounts on the same platform can be used to deanonymise users or to identify sockpuppet clusters \cite{MitigateAbuseAV_2022}. Being able to detect accounts held by the same person allows for better mitigation strategies as well as an understanding of manipulative behaviour. On Wikipedia, identifying sockpuppets helps prevent interest groups from influencing the narrative 
\cite{sockpuppetswikipedia,raszewski-de-kock-2025-detecting}. 
Writing style can also be used to link users across platforms \cite{StyleLink_2025}, for example, to deanonymise accounts on dark web forums by linking them to clear-web accounts \cite{ALightinTheDarkWeb}.


\paragraph{Topic and other biases}
Previous work has already explored the importance of one important confounding factor on authorship classification performance: topic. Features used in stylometry have long been categorised in terms of how much style as opposed to content they appear to encode \cite{writeprints, stamatatos2009}. The actual impact of topic has been studied in a variety of ways. For example, \citet{MitigateAbuseAV_2022} use feature analysis to ensure the most important features are not purely topic related. \citet{topicOrStyle} perform an ablation study to highlight the importance of content-encoding features for stylometry. Finally, \citet{stylisticfeatiuresinembeddings} tested the transformer-based LUAR-Model~\cite{LUAR} for its capacity to encode writing style distinct from content. All three show that topic is an important confounding factor to consider. 

Other work focuses on influences on writing style, such as genre/domain \cite{bartelds2019improving,barlas2020cross} or the dependency on artefacts in a given corpus \cite{BIASAV, BenchmarkBias}. Most of these papers do not frame the influence of topic, genre/domain or corpus artefacts as a question relating to bias as understood in the fairness literature, but as a threat to the validity of the technique. 

To the best of our knowledge, no work has analysed the influence of authors' demographic characteristics on authorship attribution or verification. However, there have been attempts to predict demographic characteristics such as age and gender based on text \cite{santosh2013genderage, piot2021gender}. These profiling classifiers generally overfit on the domain they were trained on \cite{age_gender_overfit}. To facilitate further investigations into the biases of authorship profiling and authorship obfuscation, \citet{emmery2024sobr} collected a dedicated Reddit dataset (SOBR) for researchers.

\section{Datasets}

While many social media datasets contain authorship information, none are apt for our task. Most existing datasets lack demographic characteristics, and those that do use unreliable proxies, such as first names, for gender. The SOBR dataset \cite{emmery2024sobr} provides texts labelled by author, gender and age, but not native language. 

To better study how a user's demographic characteristics affect authorship attribution made with Reddit data, we collect our own purpose-built dataset. For our experiments, we need a large number of users who 1) self-declare either their native language, their gender, or their age and 2) have written a sufficient number of long comments for a classification to be viable.

\begin{figure}[H]
\centering
\tikzstyle{startstop} = [rectangle, rounded corners, 
minimum width=2cm, 
minimum height=0.7cm,
text centered, 
draw=black, 
fill=red!30]

\tikzstyle{proc} = [trapezium, 
trapezium stretches=true, 
trapezium left angle=70, 
trapezium right angle=110, 
minimum width=2cm, 
minimum height=1cm, text centered, 
draw=black, fill=blue!30]

\tikzstyle{io} = [rectangle, 
minimum height=0.7cm,
minimum width=0.7cm, 
text centered, 
text width=2cm, 
draw=black, 
fill=orange!30]

\tikzstyle{arrow} = [thick,->,>=stealth]

\begin{tikzpicture}[node distance=1.5cm]

\node (start) [startstop] {reddit.com/r/};

\node (wayback) [proc, right of=start, xshift=2.5cm] {1) Wayback API};

\node(thread_ids) [io, below of=start]{Thread IDs};

\node (reddit1) [proc, below of=wayback] {2) Reddit API};

\node(user_ids) [io, below of=thread_ids]{User IDs \& Flair};

\node (filter) [proc, below of=reddit1] {3) Flair-based Filter};

\node(user_ids_f) [io, below of=user_ids]{User IDs};

\node (reddit2) [proc, below of=filter] {4) Reddit API};

\node(final) [io, below of=user_ids_f]{User Post Hisotry};

\draw [arrow] (start) -- (wayback);
\draw [arrow] (wayback) -- (thread_ids);
\draw [arrow] (thread_ids) -- (reddit1);
\draw [arrow] (reddit1) -- (user_ids);
\draw [arrow] (user_ids) -- (filter);
\draw [arrow] (filter) -- (user_ids_f);
\draw [arrow] (user_ids_f) -- (reddit2);
\draw [arrow] (reddit2) -- (final);

\end{tikzpicture}

\captionof{figure}{Diagram of our data collection methodology. The left column indicates the input and output of each step, and the right column indicates the process used at each step. }

\label{fig:flowchart_collection}
\end{figure}
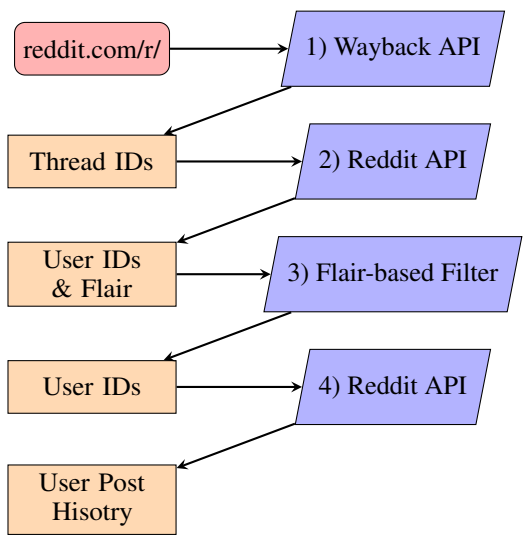

\subsection{Data Collection}
\label{sec:data}

To gather such a dataset at scale, we developed a novel data collection methodology that combines the Wayback API \url{archive.org}\footnote{\url{https://archive.org/help/wayback_api.php}} and the Reddit API\footnote{\url{https://www.reddit.com/dev/api/}}. 

The Reddit API only returns the 1k most recent threads for a particular subreddit. However, if queried directly with a \textit{thread\_id} that is older than this imposed limit, the Reddit API returns the requested resource. This is the mechanism we use to collect more data by using the Wayback API.

Our collection methodology is illustrated in Figure~\ref{fig:flowchart_collection}.
At a high level, we collect data by first finding subreddits where users publicly share their demographic information in their user \textit{flair}, a subreddit-specific tag that is displayed next to the username.
Then, we collect as many \textit{thread\_ids} from posts in these subreddits using the Wayback API. Using the Reddit API, we subsequently collect all \textit{user\_ids} of users who participated in those threads, including their relevant \textit{flair} (e.g., age), and finally, we collect the post history of each user whose flair identified them as being a member of one of the demographic groups we wanted to study.
Using this methodology, we collect three different datasets. 

\paragraph{Native Language}
The first dataset consists of users of r/languagelearning, a subreddit dedicated to language learning.  We reduce the rather complex flairs available in this community into their native language and, if available, their English skill. In total, we were able to collect 411 authors suitable for our experiments, 161 of whom indicate being native English speakers and 250 indicate being non-native English speakers. The majority of non-native English speakers indicate a high level of proficiency in English. 
\paragraph{Gender}
The Gender dataset is collected from five advice subreddits where the goal is to ask questions to people of a specific gender (r/askmen, r/askwomen, r/asktransgender, r/askmenover30, r/askwomenover30), two subreddits focused on transgender issues (r/mtf, r/ftm), and two others (r/twoXindia, r/sexover30). Based on the flair conventions in the largest of these subreddits at the moment of collection, we categorised people into three categories: woman, man, and trans*. There is no way of knowing whether somebody choosing a flair was describing their sex or their gender, especially because of varying subreddit conventions; thus, what we call gender here is a proxy variable that could indicate both.
To avoid assigning users labels that they would not use to self-describe if based on their flairs, multiple labels could be assigned using our parsing; we remove the users from our dataset.
A total of 6,982 users met our data requirement, of which 1,149 describe themself as "trans*", 2,842 self-identify as "man", and 2,991 self-identify as "woman". 

\paragraph{Age}
 We collected the third dataset from seven subreddits focused on generation-related topics (r/generationology, r/BabyBoomers, r/GenX,r/Xennials, r/Milenials, r/Milennials, r/Zillenials). We do not assign a user a generation based only on their participation in that generational subreddit, since users are free to comment in subreddits of different generations. We instead use flair to assign users to a generation, since in all but the Millennial subreddits, the flair indicates the user's year of birth. Our dataset has fewer Millennial users due to the more ambiguous flair conventions found on those subreddits.  We run our experiments on a generational basis. We compare users with a year of birth between 1968 and 1972, associated with the Generation X (GenX), to users born between 1999 and 2003, attributed to the Zoomer (GenZ) generation. We chose these two generational subsets because we had more than 160 users for each generation within this timespan, and all inter-generational age differences are larger than the intra-generational ones.

\paragraph{Data Preprocessing}
We perform all our experiments at a comment-by-comment level. 
Before using the comments, we strip them of emojis, Reddit quotations and URLs. Furthermore, we determine the language they are written in using polyglot \cite{polyglot} and only retain comments written in English.
In all our experiments, we train with between 4,950 and 5,050 words worth of comments that are at least 128 words long. We based this decision on tests with shallow classifiers, where our goal was ``good-enough'' classification with the least training data available. 
We use 10 comments of a minimal length of 128 words for testing. To mimic a real-world setting, we use a user's oldest texts for training and use their newer texts for testing.

\section{Classification Scheme}
\label{shallow_clas_feat}
We evaluated a variety of authorship attribution schemes to determine the best-performing model (see Appendix). A reduced version of the \textit{writeprints} features described by~\citet{writeprints} in combination with Logistic Regression (logR) achieved the best results. The results displayed in the following three experiments use this combination. However, to corroborate our results, we also ran the experiment with a logistic Regression in combination with normalised character-n-grams. We noted no difference in the result, except for a decrease in classifier performance. Furthermore, we tested different machine learning algorithms in combination with the two aforementioned feature sets on the native language dataset, again with the same overarching results presented in the following sections. 

\section{Impact of Demographic Characteristics on Classifier Performance}
\label{Section_Experiment1}
\subsection{Experimental Setup}

We first assess how the composition of demographic characteristics in the suspect set impacts the performance of authorship attribution. 

For this, we vary what we refer to as \textit{suspect set composition}, i.e., the proportion of users with one demographic characteristic ($dc$) compared to users with another demographic characteristic ($\neg dc$). For each $k \in \{1,...,n\}$, we run the same experimental setup for all possible suspect sets of size $n \in \{2^2, 2^3, 2^4\}$ of $k$ authors with characteristic $dc$ and $n-k$ authors with characteristic $\neg dc$. 

If the suspect set composition impacts the classification accuracy, we would expect the accuracy to increase or decrease with the number of native speakers in the suspect set. That is, if a suspect set with more, e.g., women in it, performed worse than one with more men in it, we could conclude that the proportion of women in the suspect set does impact classifier performance. If the performance does not decrease or increase, we can conclude that the suspect set composition does not affect the classifier's performance. 

\begin{figure}[H]
 \centering
 \includegraphics[width=\linewidth]{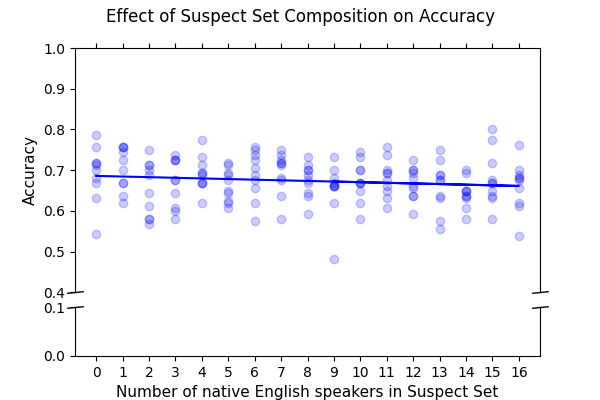}
 \captionof{figure}{The first-order polynomial predicting a classifier's accuracy based on the native language composition of the suspect set. The points represent the classification accuracy of 10 experiments for each suspect set composition (0 native speakers/16 non-native speakers to 16 native speakers/0 non-native speakers), and the line represents the linear regression result. The slope of the line is almost horizontal. This indicates visually that the performance is not influenced by this type of suspect set composition. Furthermore, the line is well-centred in the point clouds, indicating that it accurately describes the data.}
 \label{fig:sub_noR_16}
\end{figure}

We evaluate the relationship between classifier performance and suspect set composition using the first-order polynomial resulting from a linear regression. The latter is computed on the chosen performance metric (accuracy or F1) over the number of authors with characteristic $dc$ in the suspect set. This is illustrated with an example in Figure \ref{fig:sub_noR_16}. 

We use two metrics to evaluate the linear regression: the mean-squared error (MSE) to assess how well the slope fits the data and the coefficient of determination ($R^2$) to determine how well the equation explains the observed variation. The closer the former is to zero, the better the slope fits the data; the closer the latter is to one, the better the suspect set composition explains the observed variability in the performance metric. For a given suspect set composition, we generate 10 different suspect sets with no overlap in authors.

\subsubsection{Results}

\begin{table*}[htb]
\begin{center}
\begin{tabular}{p{0.5cm}lllrrrrrr}\toprule

& \multicolumn{4}{c}{\textbf{Experiment Setup}} & \multicolumn{4}{c}{\textbf{Regression Metrics}} \\
\cmidrule(lr){2-5}
\cmidrule(ll){6-9}
 & \boldmath{$dc$} & \boldmath{$\neg dc$} &  & \textbf{\# Users} & \textbf{y-intercept }& \textbf{slope}& \textbf{mse} & \boldmath{$R^2$} \\
&&&&&&&&\\
\cmidrule(lr){1-1}
\cmidrule(lr){2-5}
\cmidrule(ll){6-9}

\multirow{3}{*}{\rotatebox{90}{\parbox{0.75cm}{Age}}} & GenX & GenZ & & 4 & 0.83 & -0.02 & 0.01 & 0.05 \\
 & GenX & GenZ & & 8 & 0.82 & -0.01 & 0 & 0.14 \\
 & GenX & GenZ & & 16 & 0.76 & -0.01 & 0 & 0.19 \\ 
&&&&&&&&\\
\cmidrule(lr){1-1}
\cmidrule(lr){2-5}
\cmidrule(ll){6-9}

\multirow{9}{*}{\rotatebox{90}{\parbox{1.5cm}{Gender}}} & F & T & & 4 & 0.89 & -0.01 & 0.01 & 0.03 \\
 & F & T & & 8 & 0.79 & 0 & 0 & 0 \\
 & F & T & & 16 & 0.72 & 0 & 0 & 0.04 \\ 
\cmidrule(lr){2-5}
\cmidrule(ll){6-9}
 & M & F & & 4 & 0.8 & 0.02 & 0.01 & 0.1 \\
 & M & F & & 8 & 0.78 & 0 & 0 & 0 \\
 & M & F & & 16 & 0.7 & 0 & 0 & 0 \\ 
 \cmidrule(lr){2-5}
\cmidrule(ll){6-9}
 & M & T & & 4 & 0.82 & 0.02 & 0.01 & 0.07 \\
 & M & T & & 8 & 0.8 & 0 & 0.01 & 0.01 \\
 & M & T & & 16 & 0.73 & 0 & 0 & 0.04 \\ 
 &&& \multicolumn{2}{l}{\boldmath{$\neg$}\textbf{N Lang. Selection}} &&&&\\
\cmidrule(lr){1-1}
\cmidrule(lr){2-5}
\cmidrule(ll){6-9}

\multirow{6}{*}{\rotatebox{90}{\parbox{2cm}{Native Lang.}}} & N & $\neg{N}$ & Random NL & 4 & 0.86 & -0.01 & 0.01 & 0.02 \\
 & N & $\neg{N}$ & Random NL & 8 & 0.76 & 0 & 0.01 & 0 \\
 & N & $\neg{N}$ & Random NL & 16 & 0.69 & 0 & 0 & 0.02 \\ 
 \cmidrule(lr){2-5}
\cmidrule(ll){6-9}
 & N & $\neg{N}$ & Shared NL & 4 & 0.84 & 0 & 0.01 & 0 \\
 & N & $\neg{N}$ & Shared NL & 8 & 0.77 & 0 & 0.01 & 0.03 \\
 & N & $\neg{N}$ & Shared NL & 16 & 0.67 & 0 & 0 & 0.01 \\\bottomrule

\end{tabular}
 \end{center}
\caption{This table describes the first-order polynomial resulting from a linear regression, where, based on the number of $dc$ in the suspect set ($x$), we predict the accuracy ($y$).The compared demographic factors are listed in the columns $dc$ and $\neg{dc}$. The y-intercept (y-int) describes the mean value for the accuracy [$y$], whereas the slope describes the steepness of the line. The regression quality is evaluated using the mean squared error (mse) and $R^{2}$. The \# Users column indicates the number of authors in the suspect set. The $\neg N \ Lang. Selection$ column indicates whether the non-native authors in the experimental setup necessarily share a native language \textit{Shared NL} or not \textit{Random NL}. The slopes of every experiment are near zero, indicating that the suspect set composition has no impact on the classifier's performance.}
\label{tab:supertable_reg}
\end{table*}

Table~\ref{tab:supertable_reg} shows the slopes, y-intercepts, and evaluation metrics describing the first-order polynomials for different suspect set sizes and different demographic factors. 
The slope, the $mse$, as well as the $R^2$, are all close to zero. This indicates that there is little to no influence of age, gender, or native language on classification, as the regression line is consistently horizontal. Furthermore, this horizontal line fits the data well, but there are other variables influencing the variability of the observed data. 

\begin{figure*}[t]
 \centering
 \includegraphics[width=\linewidth]{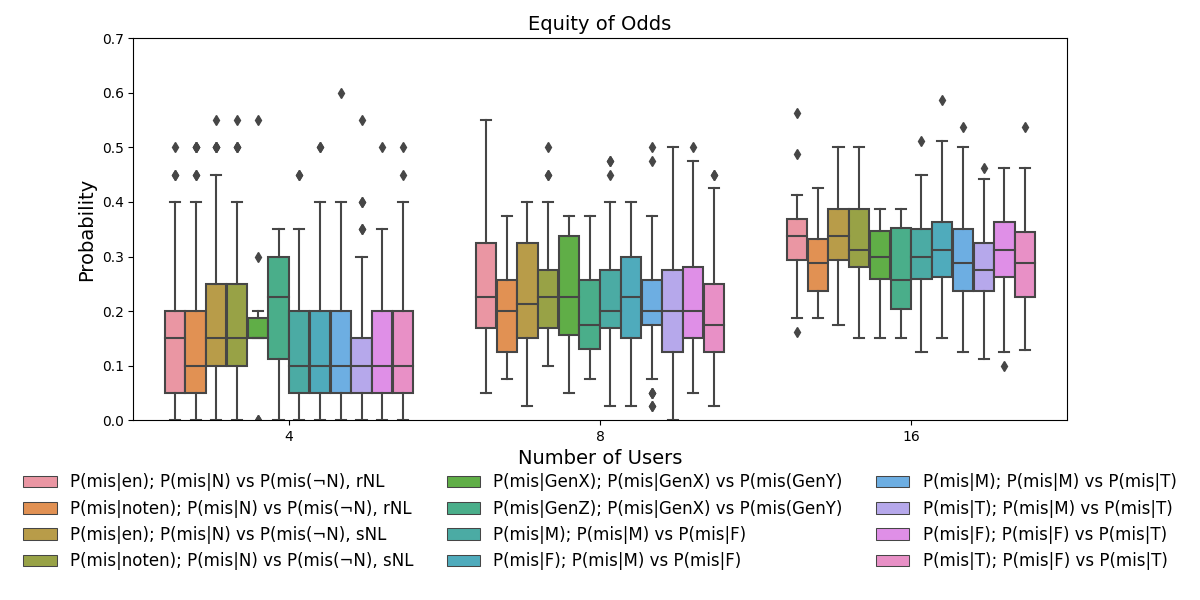}
 \caption{Boxplots showing the probability distributions of being misclassified depending on your demographic characteristic. This figure recombines multiple experiments; the different shades are labelled by the probability followed by the experimental setup. For the native language experiment, we run two experiments: \textit{sNL}, where non-native authors share a native language, and \textit{rNL}, where the non-native authors are randomly chosen.}
 \label{fig:superfigure}
\end{figure*}

\section{Equity of Odds}
\label{Section_Experiment2}
\subsection{Experiment Setup}
In this experiment, we determine how a user's demographic characteristic influences the probability that their text is misclassified. The fairness literature refers to this metric as \textit{equity of odds} or \textit{predictive parity}~\cite{ch3fairml}. Fairness is achieved for this metric if all groups of a population have the same probability of having a positive or negative classification outcome~\cite{dieterich2016compas}. 

In our case, a positive outcome is a correct classification, and a negative outcome is a misclassification.
We test the equality of odds by comparing the probability distributions of a text of an author being misclassified, given their demographic characteristic, which is noted as: $P(mis | dc )=P(mis\cap dc)/P(dc)$. In this notation, the fairness criteria translates to $P(mis | dc )\approx P(mis|\neg dc)$. We evaluate whether these distributions are statistically significantly different using a two-sided Wilcoxon Ranksum Test as implemented by \citet{2020SciPy-NMeth}. 

The null hypothesis of this test is that the observed samples, in our case, the probabilities, are drawn from the same distribution. If the hypothesis is rejected with a given p-value, it is a clear indication that the underlying distributions are not the same for the said threshold. However, if the Null hypothesis is not rejected, this only indicates that the underlying probability distribution might be the same, not that it necessarily is. We choose this non-parametric test because it makes minimal assumptions about the underlying probability distribution; for example, it does not require a normal distribution, and it works with small and differently sized sample groups. We tested whether the probabilities we compared are normally distributed using the Shapiro-Wilk test, and while some sample sets passed the test, not all of them did; thus, using a statistical test that requires normality is not an option. The Wilcoxon Rank Sum Test requires that the compared sets of samples are independent, which we can guarantee. We test this in a setting where the prior probability of the assignment of a demographic characteristic is random chance.

\subsection{Results}
The probability distributions of being misclassified given a specific demographic characteristic $dc$ are largely equivalent between $dc$ and $\neg dc$. This can be seen in Figure~\ref{fig:superfigure}, where the results of the experiments for gender, native language and age are displayed. The p-values from the Wilcoxson Ranksum comparing a $P(mis|dc)$ to $P(mis|\neg dc)$ are above the threshold of 0.05. Thus, the null hypothesis of the samples originating from the same distribution cannot be rejected. This indicates that in our datasets, an author's demographic characteristics do not influence the rate at which their text is misclassified.

\section{Forced-Misclassification}
\label{Section_Experiment3}
In previous experiments, we only considered misclassifications as they occur in a closed-world setting. In these experiments, errors are relatively rare, and those that do occur are mostly explained by atypical deviation in the writing style of the author. For example, a user might employ a formal tone on r/AskHistorians, but a more informal tone on r/aww, and a more combative tone on r/Politics. 

The forced misclassification setting highlights the expected behaviour of supervised machine learning models: test samples that are close to training samples in a way that is distinct from other authors are correctly classified, while those that deviate from the training data are more likely to be misclassified. To illustrate this with an example, we chose at random an experimental setup from the previous sections (Male/Female, 4 users) and measured the cosine distance from the mean training vector to each of the testing samples. The closer the cosine distance is to 0, the more similar the two compared vectors are.
Unsurprisingly, we find that the misclassified samples have an average distance of 0.95 from the average training vector, and the correctly classified samples are much closer, with an average distance of~0.84. 

\begin{figure*}[t]
 \centering
 \includegraphics[width=\linewidth]{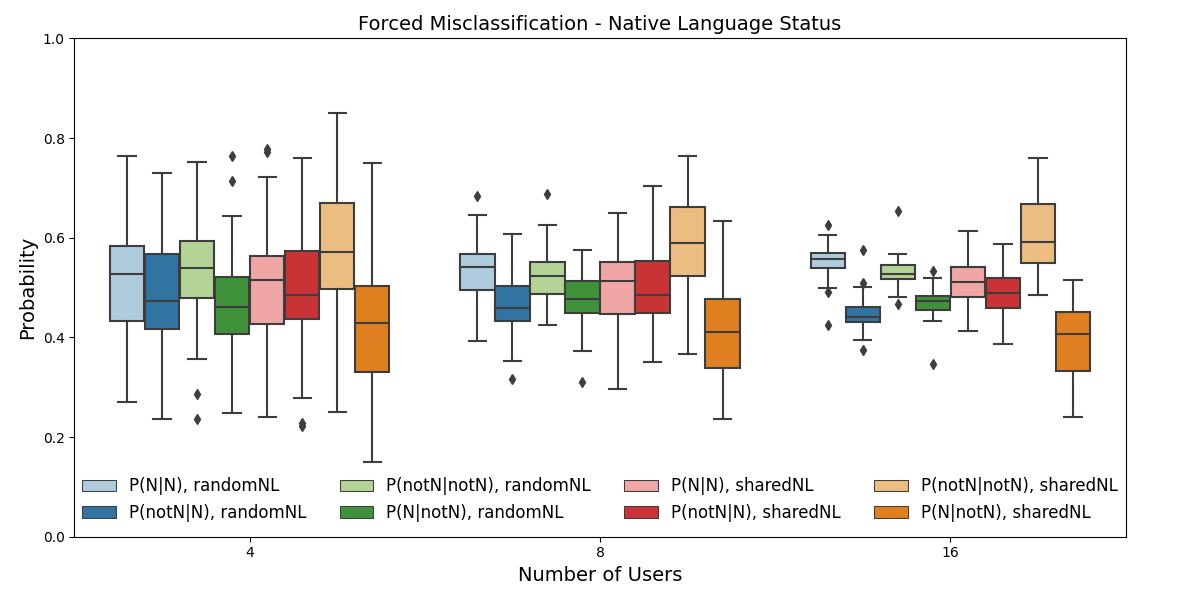}
 \caption{Boxplots displaying the probability distribution of intra- and inter-native language assignment for different native languages. $N$ stands for native English speaker, $\neg N$ stands for non-native English speaker. In the experiment setting $randomNL$, non-native authors are randomly chosen, as opposed to the experiment setting $sharedNL$, where non-native authors have a common native language.}
 \label{fig:leave_out_NL}
\end{figure*}

Because we want to study the influence of confounding demographic factors on mistakes, we eliminate the influence of a user's deviation from the following experiment. To this end, we perform the following experiment in a setting we refer to as \textit{forced misclassification}. In this setting, the true author of the tested comment is not in the training data of the classifier. The forced-misclassification setting differs slightly from the open-world setting found in the literature. We do not test the classifier on some labels that are in the training set and some that are not. We want to establish if users who share a demographic characteristic with the true author write in a similar style, thus systematically get mistaken for each other. To gain insight into the influence of the demographic characteristic on the classification in this setting, we compare the following probability distributions:
The probability that an author who shares the demographic characteristic with the true author gets assigned the text $P(dc|dc)$. We refer to this as an intra-status assignment.
The probability that an author with a different demographic characteristic is assigned the text, $P(dc|\neg dc)$, we call this an inter-status assignment. We test this in a setting where the prior probability of the assignment of a demographic characteristic is random chance.

\subsection{Native Language}
Figure~\ref{fig:leave_out_NL} displays the probability distributions for 4, 8, and 16 classes. 
In the experiment setting where the non-native authors are randomly chosen (the green and blue boxplots in Figure~\ref{fig:leave_out_NL} and the Wilcoxson Rank Test results in Table~\ref{tab:randomnl_pvalue}), intra-status assignments of text occur at a higher rate for non-native speakers than inter-status assignments. For native speakers, intra-status assignments occur at higher rates for a suspect set size of 8 and 16 users. For 4 users, the hypothesis that intra- and inter-status assignments occur at the same rate cannot be rejected.
The rates of intra-status assignments for native and non-native speakers for a suspect set size of 4 or 8 users are similar.
However, for 16 authors, the distributions are statistically significantly different. The same is true for the rates of inter-status assignment.

For the setup where non-native authors share a native language (the orange and red boxplots in Figure~\ref{fig:leave_out_NL} and the Wilcoxson Rank Test results in Table~\ref{tab:shared_nl}), the results are different. The rates of intra-status assignments for texts of native and non-native speakers are statistically significantly different, independent of suspect set size. The same is true for inter-status assignments and the comparison between intra- and inter-status assignments for texts of non-native authors. For native authors, intra- and inter-status assignments for texts occur at different rates for a suspect set of 16 authors. For a suspect set size of 4 or 8, the Wilcoxon Rank Sum Test does not reject the null hypothesis, thus the probability distributions might be the same.

In brief, for sock-puppet detection on a subreddit like r/europe, where it is known that non-native authors have a diverse set of native languages, the classifier would behave similarly for native and non-native authors. However, on a subreddit where it is assumed that non-native authors all share a native language, the tendency of the classifier to make intra-status assignments is higher for non-native authors as compared to native authors.

\begin{table*}[htb]
    \centering

    \begin{subtable}{0.48\textwidth}
        \begin{tabular}{llllll}\toprule
         \textbf{} & 
         \multicolumn{2}{c}{\textbf{Group}} & 
         \multicolumn{2}{c}{\textbf{p-value}} \\
        \cmidrule(lr){2-3}
        \cmidrule(ll){4-5}
        \textbf{Users} & 1 & 2 & \textless{}0.05 & \textless{}0.01 \\
        \cmidrule(lr){1-1}
        \cmidrule(lr){2-2}
        \cmidrule(lr){3-3}
        \cmidrule(lr){4-4}
        \cmidrule(ll){5-5}
        4 &\textbf{$P(\neg N|\neg N)$} & \textbf{$P(N|\neg N)$} & \textbf{True} & \textbf{True} \\
        4 & $P(N|N)$ & $P(\neg N|N)$ & \textbf{True} & False \\
        4 & $P(N|\neg N)$ & $P(\neg N|N)$ & False & False \\
        4 & $P(N|N)$ & $P(\neg N|\neg N)$ & False & False \\
        8 & \textbf{$P(\neg N|\neg N)$} & \textbf{$P(N|\neg N)$ } & \textbf{True} & \textbf{True} \\
        8 & \textbf{$P(N|N)$ } & \textbf{$P(\neg N|N)$} & \textbf{True} & \textbf{True} \\
        8 & $P(N|\neg N)$ & $P(\neg N|N)$ & False & False \\
        8 & $P(N|N)$ & $P(\neg N|\neg N)$ & False & False \\
        16 & \textbf{$P(\neg N|\neg N)$} & \textbf{$P(N|\neg N)$} & \textbf{True} & \textbf{True} \\
        16 & \textbf{$P(N|N)$ } & \textbf{$P(\neg N|N)$ } & \textbf{True} & \textbf{True} \\
        16 & \textbf{$P(N|\neg N)$} & \textbf{$P(\neg N|N)$} & \textbf{True} & \textbf{True} \\
        16 &\textbf{ $P(N|N)$} & \textbf{$P(\neg N|\neg N)$} & \textbf{True} &\textbf{True}\\\bottomrule
         \end{tabular}
         \caption{\textbf{Native Language}: Non-English-native suspects do not necessarily have the same native language}
         \label{tab:randomnl_pvalue}
    \end{subtable}\hfill
    \begin{subtable}{0.48\textwidth}
    \centering
    \begin{tabular}{llllll}\toprule
         \textbf{} & 
         \multicolumn{2}{c}{\textbf{Group}} & 
         \multicolumn{2}{c}{\textbf{p-value}} \\
        \cmidrule(lr){2-3}
        \cmidrule(ll){4-5}
        \textbf{Users} & 1 & 2 & \textless{}0.05 & \textless{}0.01 \\
        \cmidrule(lr){1-1}
        \cmidrule(lr){2-2}
        \cmidrule(lr){3-3}
        \cmidrule(lr){4-4}
        \cmidrule(ll){5-5}
        4 & $P(\neg N|\neg N)$ & $P(N|\neg N)$ & \textbf{True} & \textbf{True} \\
        4 & $P(N|N)$ & $P(\neg N|N)$ & False & False \\
        4 & $P(N|\neg N)$ & $P(\neg N|N)$ & \textbf{True} & \textbf{True} \\
        4 & $P(N|N)$ & $P(\neg N|\neg N)$ & \textbf{True} & \textbf{True} \\
        8 & $P(\neg N|\neg N)$ & $P(N|\neg N)$ & \textbf{True} & \textbf{True} \\
        8 & $P(N|N)$ & $P(\neg N|N)$ & False & False \\
        8 & $P(N|\neg N)$ & $P(\neg N|N)$ & \textbf{True} & \textbf{True} \\
        8 & $P(N|N)$ & $P(\neg N|\neg N)$ & \textbf{True} & \textbf{True} \\
        16 & $P(\neg N|\neg N)$ & $P(N|\neg N)$ & \textbf{True} & \textbf{True} \\
        16 & $P(N|N)$ & $P(\neg N|N)$ & False & False \\
        16 & $P(N|\neg N)$ & $P(\neg N|N)$ & \textbf{True} & \textbf{True} \\
        16 & $P(N|N)$ & $P(\neg N|\neg N)$ & \textbf{True} & \textbf{True}\\\bottomrule
        \end{tabular}
        \caption{\textbf{Native Language}: Non-English-native suspects all share the same native language}
        \label{tab:shared_nl}
    \end{subtable}

    \begin{subtable}{0.48\textwidth}
    \centering
     \centering
         \begin{tabular}{llllll}\\\toprule
         
         \textbf{} & 
         \multicolumn{2}{c}{\textbf{Group}} & 
         \multicolumn{2}{c}{\textbf{p-value}} \\
        \cmidrule(lr){2-3}
        \cmidrule(ll){4-5}
        \textbf{Users} & 1 & 2 & \textless{}0.05 & \textless{}0.01 \\
        \cmidrule(lr){1-1}
        \cmidrule(lr){2-2}
        \cmidrule(lr){3-3}
        \cmidrule(lr){4-4}
        \cmidrule(ll){5-5}
        4 & $P(M|M)$ & $P(F|F)$ & False & False \\ 
        4 & $P(F|M)$ & $P(M|F)$ & False & False \\
        \textbf{4} & \textbf{$P(M|F)$} & \textbf{$P(F|F)$} & \textbf{True} & \textbf{True} \\ 
        \textbf{4} & \textbf{$P(F|M)$} & \textbf{$P(M|M)$} & \textbf{True} & \textbf{True} \\
        8 & $P(M|M)$ & $P(F|F)$ & False & False \\ 
        8 & $P(F|M)$ & $P(M|F)$ & False & False \\
        \textbf{8} & \textbf{$P(M|F)$} & \textbf{$P(F|F)$} & \textbf{True} & \textbf{True} \\ 
        \textbf{8} & \textbf{$P(F|M)$} & \textbf{$P(M|M)$} & \textbf{True} & \textbf{True} \\
        16 & $P(M|M)$ & $P(F|F)$ & False & False \\ 
        16 & $P(F|M)$ & $P(M|F)$ & False & False \\
        \textbf{16} & \textbf{$P(M|F)$} & \textbf{$P(F|F)$} & \textbf{True} & False \\ 
        \textbf{16} & \textbf{$P(F|M)$} & \textbf{$P(M|M)$} & \textbf{True} & False \\\bottomrule
        \end{tabular}
        \caption{\textbf{Gender}}
        \label{pvalue_table}
    \end{subtable}\hfill
    \begin{subtable}{0.48\textwidth}
    \centering
    \begin{tabular}{llllll}\toprule
         \textbf{} & 
         \multicolumn{2}{c}{\textbf{Group}} & 
         \multicolumn{2}{c}{\textbf{p-value}} \\
        \cmidrule(lr){2-3}
        \cmidrule(ll){4-5}
        \textbf{Users} & 1 & 2 & \textless{}0.05 & \textless{}0.01 \\
        \cmidrule(lr){1-1}
        \cmidrule(lr){2-2}
        \cmidrule(lr){3-3}
        \cmidrule(lr){4-4}
        \cmidrule(ll){5-5}
        4 & $P(X|X)$ &$P(Z|X)$ & \textbf{True} & False \\
        4 &$P(Z|Z)$ &$P(X|Z)$ & False & False \\
        4 &$P(Z|Z)$ &$P(X|X)$ & False & False \\
        4 &$P(Z|X)$ &$P(X|Z)$ & False & False \\
        8 &$P(X|X)$ &$P(Z|X)$ & \textbf{True} & \textbf{True} \\
        8 &$P(Z|Z)$ &$P(X|Z)$ & \textbf{True} & False \\
        8 &$P(Z|Z)$ &$P(X|X)$ & False & False \\
        8 &$P(Z|X)$ &$P(X|Z)$ & False & False \\
        16 &$P(X|X)$ &$P(Z|X)$ & \textbf{True} & \textbf{True} \\
        16 &$P(Z|Z)$ &$P(X|Z)$ & \textbf{True} & \textbf{True} \\
        16 &$P(Z|Z)$ &$P(X|X)$ & False & False \\
        16 &$P(Z|X)$ &$P(X|Z)$ & False & False \\\bottomrule
        \end{tabular}
        \caption{\textbf{Generation}: X = GenX, Z = GenZ}
        \label{pvalue_table_gen}
    \end{subtable}

    \caption{Summary of the Wilcoxson Rank Sum test results. The Group column lists the compared probability distributions. N stands for native English speakers, $\neg N$ stands for non-native English speakers. The p-value columns describe whether the p-values are below a given threshold. Statistically significant differences are put in bold. The \textit{Users} column lists the number of authors in the training set.}
    
\end{table*}

\subsection{Gender}
Intra-status assignments occur at similar rates for different genders. The same is true for inter-status assignments. This can be seen in Figure~\ref{fig:leave_out_gender} by comparing $P(M|M)$ to $P(F|F)$ and $P(F|M)$ to $P(F|M)$. This is confirmed by the Wilcoxon Rank Sum results, where for these comparisons the Null-hypothesis is not rejected (Table~\ref{pvalue_table}).

For both women and men, intra-status assignments occur at statistically significantly higher rates than inter-status assignments. I.e., text from a woman has a higher likelihood of being assigned to another woman rather than to a man. However, this only holds for a suspect set size of 16 if we use a less strict cut-off for statistical significance of p-value~$< 0.05$ (see Table~\ref{pvalue_table}). 

\subsection{Age}
\begin{figure*}
\centering
\begin{minipage}{.48\textwidth}
  \centering
  \includegraphics[width=\linewidth]{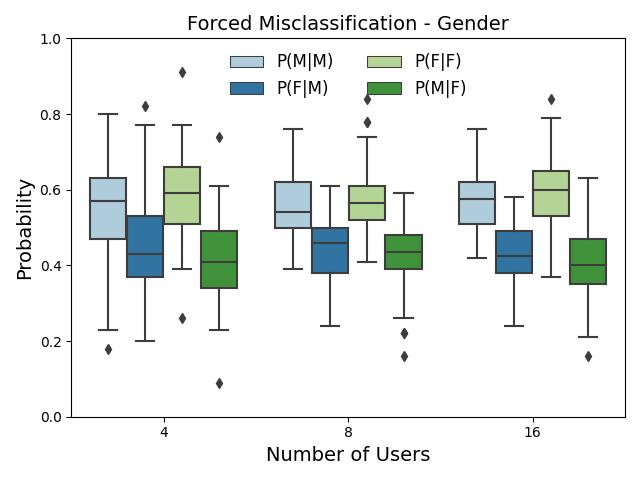}
  \captionof{figure}{The probability distributions of intra- and inter-gender assignments for different genders.}
  \label{fig:leave_out_gender}
\end{minipage}%
\hfill
\begin{minipage}{.48\textwidth}
  \centering
  \includegraphics[width=\linewidth]{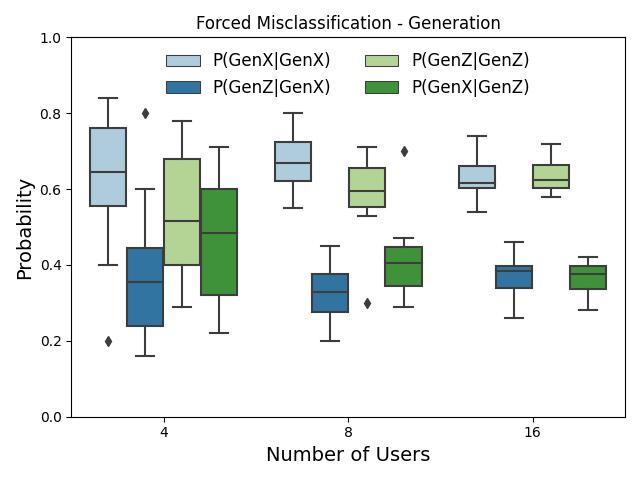}
  \captionof{figure}{Boxplots of displaying the probability distributions of intra- and inter-generation assignment.}
  \label{fig:leave_out_gen}
\end{minipage}
\end{figure*}

Intra-status assignments occur at similar rates for different generations. The same is true for inter-status assignments. This can be seen in Figure~\ref{fig:leave_out_gen} and is confirmed by the Wilcoxon Rank Sum results (see Table~\ref{pvalue_table_gen}).

When comparing the intra- and inter-generational assignments, the results differ depending on the number of users (see Table~\ref{pvalue_table_gen}).
For a suspect set size of 16 authors, the intra- and inter-generational assignments occur at different rates. The text has a statistically significant tendency to be assigned to an author of the same generation.
For a suspect set size of 8 authors, the same is true, except with the caveat that for authors belonging to GenZ, this is only the case with the less strict significance cut-off p-value $<0.05$.
In contrast, for the suspect set size of 4, this pattern of statistical significance is no longer given.

\section{Discussion}
In a closed-world setting, our results do not indicate any influence of demographic factors on authorship attribution. 
In our first experiment, we studied the impact of the demographic composition of the training data on classifier performance. Our results show that the diversity of the training set with regard to gender, age or native language does not impact the performance of authorship attribution. 

Therefore, when building an authorship attribution-based classifier, for example, to link sockpuppet accounts on Reddit, there is no performance incentive to pay attention to the demographic makeup of the authors in the training set. It works equally well regardless of the demographic makeup of the suspect set. However, these results also indicate that demographic traits do not provide any natural cover for any users. For example, native English speakers are just as easy to deanonymise as non-native English speakers.

In the second experiment, we study the equity of odds: whether a user's demographic trait affects the rate at which their text is misclassified. Our results indicate that the weight of misclassification is not disproportionately carried by one group. In other words, no particular demographic group that we studied would be disproportionately impacted by model errors in automated moderation using authorship attribution \textit{in the closed-world setting}. At the same time, these results also indicate that such demographic factors cannot aid in building defences against authorship attribution attacks on forms of legitimate anonymity (e.g., whistleblowing). They also signal that everyone is equally vulnerable to an attack against their anonymity.

In our third experiment, we studied whether the demographic characteristic of the true author influences to whom a text is misattributed. Because the mistakes made in the closed world setting are influenced by an author deviating from their own normal style, we study mistakes made in the \textit(forced misclassification) setting. We present the classifier with text from users not in the training data, then measure at what rate the text gets attributed to users who share the demographic trait of interest with the true author. In this setting, we observe a tendency to assign text to users who share demographic characteristics with the true author of the text. For non-native authors, this is dependent on the experimental setup; if in the training data users have a diverse set of native languages, this effect is less pronounced than if all non-native authors share a native language. 
These results show that even though the classifier performance is not impacted by demographic characteristics and the burden of errors is not on a particular subgroup \textit{in the closed-world setting}, there still can be bias in the model. Because this experiment relies on forcing misclassifications, it more closely mimics an open-world setting. This setting is much more realistic in the context of social networks, as the suspect set is rarely known. 

In the context of using authorship attribution, this means that false positives can be unequally distributed among users who share a demographic trait with the malicious account. 

Overall, our tests indicate that authorship attribution is fair and robust under closed-world settings with regard to the tested demographic characteristics. This may be because text misclassified in a closed-world setting is, by definition, a deviation from the author's normal writing style. Therefore, their demographic trait might not have as much of an impact on the classification. However, when we force the classifier to make an error by removing the true author from the training set, we are able to evaluate all samples, regardless of their closeness to the author's average writing style. In this setting, we find that there is a statistically significant influence of the demographic characteristic. 

\subsection{Threats to validity}
\paragraph{Self-reported labels and proxy labels}

Self-reported labels are useful because they allow a user to self-identify their demographics.  However, they come with two notable drawbacks. First, nothing prevents users from lying in their flairs, thus poisoning our dataset. Second, as is made clear in our gender label, some terms can have multiple meanings. For instance, words some people use for gender are used by others to describe their sex. When we see the flairs, we cannot be sure which way the user meant the flair; thus, we are only able to measure a proxy variable. 

\paragraph{Generalizability}
Reddit's user demographic is not representative of the population at large, and the text written on Reddit follows platform and subreddit-specific conventions. While we control for one demographic characteristic at a time, the other demographic characteristics in our experiments are influenced by Reddit´s user base. Furthermore, we only perform our experiments on English text.

\paragraph{Selection Bias}
Our experimental setup requires authors to have at least 6,280 words worth of comments written in English. This criterion was met by users who are active in a variety of subreddits over a relatively long period of time (on average, one to two years). Thus, our suspect sets are mainly composed of long-term Reddit users. We also assume the main account associated with a puppet-master would also be rather prolific, and thus meet our text requirement and appear in our dataset. 

Especially when it comes to native language, our preprocessing has an impact on the type of users selected in our dataset. The minimum amount of text required selected for users with a high English language skill.

\paragraph{Intentional Changes of Writing Style}
If the author uses language correction tools, translation services or writing from generative AI, we assume that they do so consistently. Thus, their ``measured writing style'' can be impacted by this tool usage, but because the use of the tools is consistently impacting their ``writing style'', this should not impact our classification greatly. If a user decides to mimic someone else's writing style, adopts very specific in-group speak in certain circumstances, or otherwise greatly deviates from their personal norm, we do not have any mechanism in place to detect this behaviour change. We assume that all comments made from a specific user account are equally representative of the user's writing style. Furthermore, we assume that Reddit accounts are not shared, or if they are, that the demographic attribute mentioned in the flair applies to every person using the account. 

\paragraph{Intersectionality}
The datasets we collected have a small overlap in users. However, the distribution of demographic characteristics of this intersection does not allow for experiments analysing the influence of multiple characteristics at a time. Based on our results, we cannot make any statements about which demographic characteristic has a stronger influence or what their combined influence looks like.

\bibliography{Languages}
\bibliographystyle{unsrtnat}

\appendix
\section{Data Collection}
\subsubsection{Detailed Explanation}\hfill

Figure~\ref{fig:flowchart_collection} illustrates our data collection methodology.

\step{1}{Query the Wayback API to find thread\_ids}We query the Wayback API to get all of the visible posts from a subreddit for each snapshot stored by The Wayback Machine. These were parsed to find the \texttt{thread\_ids} for the threads seen on that page.

\step{2}{Query the Reddit API to find those threads}
Using this list, we query the Reddit API to find which users commented in the thread and their \texttt{flair\_text}. 

\step{3}{Find relevant users by parsing their flair} 
From these users, we used a parser to determine, depending on the dataset, either the native language, gender or age of each user based on the indications in their flair. The flair is a free-text field filled out by the user, and, as such, there is the possibility for variety in these entries. Depending on the subreddit in question, parsing thus can be challenging.  
 
\step{4}{Query the Reddit API to collect the post histories of relevant users} 
With these usernames, we query the Reddit API again to obtain the entire post history of each user. This results in a database of usernames, their flair on the subreddit of interest, and their entire post histories. 

\paragraph{Generalizability}
Using the data collection methodology presented here, a combination of the Wayback Machine API and the Reddit API can be reused in further studies to collect a more expansive Reddit dataset. Our collection methodology could be utilised to collect Reddit data for many different types of studies in which older Reddit data or data from more than the 1k most recent posts is needed. However, this strategy has one caveat: the sample of posts acquired does not necessarily follow a specific strategy, nor is it guaranteed to be a representative slice of an activity in a given subreddit. The crawls completed by \path{archive.org} are done by multiple independent crawlers that follow different, changing crawling strategies. This results in a dataset that lacks a discernible overall collection strategy. For example, in January 2023 \path{reddit.com/r/languagelearning} was archived 10 times, in March it was archived 15 times, and in August 2023 it was archived just once. For this paper, this caveat is not important because we are only looking to find more users with flair on r/languagelearning and are not concerned with studying the content of the subreddit itself. 

\subsection{Dataset Preprocessing} 
\label{sect:preprocess}

Before we can analyse the Reddit comments we collected, we must handle the fact that they are text collected ``in the wild'' and are therefore not clean and ready for analysis. Before cleaning the dataset, we remove all users with fewer than 30 comments. This threshold was conservatively chosen with the goal of not running every comment through the expensive cleaning process. 

We developed a three-step process to clean the dataset. First, we retracted elements from the text that were not relevant to our task, including non-text elements and quoted text. In the second step, we detected the language the comment was written in and only retained English-language texts. Finally, we discarded comments that are shorter than 128 words after the pretreatment. 

\paragraph{Removing URLs, Emojis, and Cited Text}
For the first step, we removed URLs from the comments because URLs are not reflective of someone's writing style. In contrast, emojis convey meaning~\cite{holtgraves2020emoji}. That is, a writer can choose among different emojis that share a similar meaning, making this choice part of their writing style. However, we choose to remove emojis from our text to prevent information leakage, such as certain emojis taking on different meanings in different languages. As an example, the cabbage emoji can be used in French to replace the word `cute' because cabbage and cute are homophones in French. Finally, we also removed quoted text from the comments. Reddit has a specific quotation style delineated with $>$, for users to cite other posts or quote text. We remove all of these quotes to avoid mixing the `writing styles' of users. Other types of quotes were not removed.

\paragraph{Potential Problem in our text pre-processing}
While we have retracted Reddit quotes from the text, there is no guarantee that we have retracted all quoted text. Our pre-processing does not account for quotes that are not marked in the Reddit quotation style or for unmarked copy-pasted text. We also do not distinguish between edited Reddit comments and unedited Reddit comments.

\paragraph{Filtering Non-English Texts}
We then used the language detection module polyglot~\cite{al-rfou-etal-2013-polyglot} to assess the language of each comment. Polyglot offers an indication of confidence in its assessment. Only comments assessed with a confidence of 99 or 100 out of 100 were added to the dataset. Furthermore, we manually validated the assessment of 100 randomly chosen comments per language for languages that at least one of the authors was proficient enough in to label (German, English, French, Spanish and Italian). We found no false positives for this confidence level. However, this does not guarantee that our dataset is only composed of English text. In particular, comments that contain multiple languages may pass the language filtering step if the majority of the comment is in English. 

\paragraph{Minimal comment length}
In order to detect writing style, a minimum amount of text is necessary. The feature extraction results of a two-word comment, such as `Thank you', is limited and conveys little of an author's habitual linguistic choices. As most of our comments are very short, we needed to determine a minimum comment length where we could start to measure `writing style' consistently. We determined that a minimum comment length of 128 words was sufficient to achieve consistent classification results on the Native Language Dataset by testing different minimal comment lengths concerning the achieved accuracy. The longer the minimal comment length, the better the accuracy. With a minimal comment length shorter than 128 words, the achieved accuracy varies greatly depending on the authors in the suspect set. For example, for a cut-off at 64 words certain suspect set compositions only achieved random-chance performance, whereas others achieved an accuracy of 0.81 for a suspect set size of 8 authors.
Due to the limited data available, we did not choose the cut-off value where the classifier performance stopped increasing, but rather the cut-off value where the classification performance was no longer highly dependent on the suspect set composition. In other words, we did not choose the minimal length of comment in regard to achieving the highest performance. We chose the first tested comment length, where we achieved a stable classification result, e.g. for different suspect sets we achieve similar accuracies.

\paragraph{Train-Test Split}
Before classification, we split the data into training and testing sets. To mimic a real-world setting, we use a user's oldest texts for training and use their newer texts for testing. Because the users in our dataset have different time frames of activity, this cut-off point between training and testing is set on a per-user level, and we do not have one global point in time where we use all text before as training data and all text after as testing data.

\section{Dataset Access}
In compliance with the Reddit API, we cannot publicly share the collected data. If you are interested in our dataset, please contact us directly.

\section{Authorship Attribution Comparison}
\label{apx:schemes}
\paragraph{Feature-Sets}
For our experiment, we compared three different feature extraction methods:
normalized char-{1,3}-grams, a reduced version of the $writeprint$ feature set as well as sentence embeddings.
The style\_featureset (see Table~\ref{tab:features}) differs from the first implementation by \citet{writeprints}. We only consider features that are comparable between languages and apply to Reddit comments. Thus, structural features were retracted as well as features that are difficult to compare between languages, such as misspelled words. Our feature set is not language-independent, as we need to know which language the text is in before extracting the features; however, the extracted features should be comparable for Indo-European languages.
\begin{table*}
\centering
\begin{tabular}{|ll|}
\hline
\multicolumn{1}{|l|}{\textit{\textbf{Feature}}} & \textit{\textbf{\#features}} \\ \hline
\multicolumn{2}{|l|}{\textbf{Lexical features}} \\ \hline
\multicolumn{1}{|l|}{Average length of words} & 1 \\ \hline
\multicolumn{1}{|l|}{Median length of words} & 1 \\ \hline
\multicolumn{1}{|l|}{Distribution of word length} & 1 \\ \hline
\multicolumn{1}{|l|}{total number of characters} & 1 \\ \hline
\multicolumn{1}{|l|}{char-n-grams (1-grams to 3-grams)} & 1000 \\ \hline
\multicolumn{2}{|l|}{\textbf{Content-related features}} \\ \hline
\multicolumn{1}{|l|}{word-n-grams (1-grams to 3-grams)} & 1000 \\ \hline
\multicolumn{1}{|l|}{Yules-K (vocabulary richness)} & 1 \\ \hline
\multicolumn{1}{|l|}{Type-to-Token-Ratio (vocabulary richness)} & 1 \\ \hline
\multicolumn{1}{|l|}{\textbf{Syntactic Features}} & \\ \hline
\multicolumn{1}{|l|}{pos-n-grams (1-grams to -grams)} & 1000 \\ \hline
\end{tabular}
\caption{List of features composing the style\_feature set. This is a reduced version of the feature-set proposed by \citet{writeprints}} 
\label{tab:features}
\end{table*}
For the sentence embeddings, we initially wanted to use the state-of-the-art LUAR \cite{LUAR} sentence embedding. However, it is trained on Reddit data, and there seems to be an overlap between their training data and our dataset. Because the goal of our study is to measure the influence of third variables, and not data contamination, we instead chose to use a more general model where we made sure it was not fine-tuned using Reddit data\footnote{specifically this one: https://huggingface.co/sentence-transformers/multi-qa-mpnet-base-dot-v1 (last accessed: 14.09.2025)}.

\paragraph{Classifiers}
We compared RandomForest (rf), XGBoost (xgb), Support Vecor Machines (svm) as well as Logistic Regression (logR) classifiers. 

\paragraph{Amount of Training Data}
The goal was to create an authorship attribution classifier that performed relatively well and with consistent performance. 
The less training data we require, the more users were able to be included in our experiment. The longer the training data required, the better the authorship attribution model performs. We decided that 5000 words were the best trade-off, which is equivalent to roughly ten pages worth of text. We assume that the puppet-master in a sockpuppet-network would be rather prolific, and thus be able to hit our training data requirement with one of their main accounts.

\begin{figure*}[htb]
 \centering
 \includegraphics[width=\linewidth]{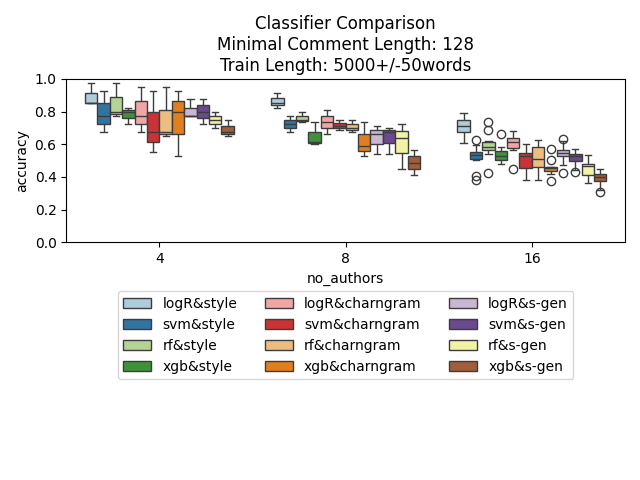}
 \caption{Comparison of classifiers, made with between 4950 and 5050 comments worth of training data. Trained and evaluated with comments that are at least 128 words long. }
 \label{fig:comparison}
\end{figure*}

\paragraph{Comparison}
Figure~\ref{fig:comparison} shows clearly that the style features in combination with a logistic regression outperform the other classification schemes. This is in accordance with the results of \citet{tyo2023valla}.

\section{Resources Required}
We ran our experiments locally on a Windows computer with 32 GB of RAM, a 128 MB Intel(R) Graphics Card, and Intel(R) Core(TM) Ultra 7 165U (1.70 GHz) Processor.

\section{Ethics}
This study only uses public data provided by the Wayback and Reddit APIs. Reddit data collected through the API has been extensively used in prior works~\cite{reddit1,reddit2,reddit3,reddit4,reddit5}. We comply with the Terms of Service of both APIs that we utilize, including respecting rate limits. In general, Reddit accounts are not tied to real-world identities, and we make no attempt to link user accounts to real people. We will also only release the dataset on request with sanitized usernames. However, for reproducibility, all code used in this project will be released prior to publication. 

\end{multicols}

\end{document}